\RequirePackage[2020-02-02]{latexrelease}
\documentclass[twocolumn]{revtex4}
\usepackage{graphicx}
\usepackage{dcolumn}
\usepackage{bm}
\usepackage{amsmath}
\usepackage{amsfonts}
\usepackage{amssymb}
\usepackage{color}
\usepackage{epstopdf}
\usepackage{float}
\providecommand{\U}[1]{\protect\rule{.1in}{.1in}}

\begin{document}
\title{Finding and improving bounds of real functions by thermodynamic arguments}
\author{Andr\'es Vallejo}
\affiliation{\begin{small} Facultad de Ingenier\'{\i}a, Universidad 
de la Rep\'ublica, Montevideo, Uruguay\end{small}}
\date{\today}
\begin{abstract}
\noindent The possibility of stating the second law of thermodynamics in terms of the increasing behaviour 
of a physical property establishes a connection between that branch of 
physics and the theory of algebraic inequalities. We use this connection to show how some well-known 
inequalities such as the standard bounds for the logarithmic function or generalizations of Bernoulli's 
inequality can be derived by thermodynamic methods. Additionally, we show that by comparing the global entropy production in processes implemented with decreasing levels of 
irreversibility but subject to the same change of state of one particular system, we can find 
progressively better bounds for the real function that represents the entropy variation of the system. As an application, some new families of bounds for the function $\log(1+x)$ are obtained by this method.

\end{abstract}

\maketitle

\section{Introduction}
In addition to the interest that algebraic inequalities present from a purely mathematical 
point of view, they are also fundamental tools in physics, economics, statistics and likely 
in any branch of science that employs mathematics in its formal description. Inequalities 
involving elementary functions appear so frequently in other disciplines that 
the problem of bounding them is an important issue.
One of the most ubiquitous elementary functions is the (natural) logarithm, 
for which the following bounds are probably well-known by the reader: 
	\begin{equation}\label{logbounds}
	\dfrac{x}{1+x}\leq \log(1+x)\leq x, \hspace{0.2cm} x>-1.
	\end{equation}
In what follows, we refer to Eq. (\ref{logbounds}) as the \textit{standard bounds} for the 
logarithmic function. From a geometric point of view, the standard bounds are 
a consequence of the concave behaviour of the function $f(x)=\log(1+x)$: since a concave 
function is upper-bounded by its tangent line at any point, $f$ is bounded by the function 
$t(x)=x$, which is its tangent line at $x=0$; the lower bound is an equivalent form of the 
upper bound, obtained by the change of variable $z+1=1/(x+1)$. Although the power and 
simplicity of the standard bounds is indisputable, in practical applications sharper bounds 
could be useful, so improvements of Eq. (\ref{logbounds}) have been obtained by several 
researchers, albeit at the expense of simplicity \cite{Love,Topsok,Chesneau}. 

In this work we address the problem of improving the bounds given by Eq. (\ref{logbounds}) 
by a method based on physical considerations.  
According to the second law 
of thermodynamics, the entropy of the Universe increases in any real physical process. 
Consequently, a symbolic expression for the entropy variation is always non-negative. 
This idea can be traced back to the 19th century, where it was first used to prove the inequality between 
the arithmetic and geometric means \cite{Tait}, a proof that has been independently rediscovered 
at least six times \cite{Sommerfeld,Cashwell,Landsberg,Tykodi,Wang,Graham}. The thermodynamic 
relevance of this particular inequality has been revealed in the context of 
studies of the efficiency of reversible heat engines operating between finite and infinite 
reservoirs.\cite{Johal}

A related concept here is the fact that the global entropy change is a measure of 
the departure of a thermodynamic process from the reversible limit. This implies that if we compare the entropy production in two processes in 
which the system undergoes the same change of state, less entropy will be produced in the implementation 
that is closer to that limit. This fact allows us to obtain a sharper bound for the entropy 
variation of the system, which, of course, is the same in both cases since entropy is a state 
function.

The method explored here not only allows students to deduce 
familiar results from an unusual starting point, but also to obtain potentially novel results 
employing accesible tools. Including this perspective while teaching these topics 
might motivate curiosity and autonomous research in students.

The outline of this paper is as follows. In Section II we obtain the standard bounds for 
the logarithmic function by performing an entropy analysis of a simple physical system undergoing
a specific process. In Section III we explain in detail the general method that allows us to improve 
the bounds for the entropy variation of the system. In Section IV we apply this method to find 
sharper bounds for the logarithmic function. Remarks and conclusions are presented in Section V. 

\section{Deriving Eq. (\ref{logbounds}) by an entropy analysis}

First, we show how the natural bounds for the logarithmic function, Eq. (\ref{logbounds}), can 
be obtained by thermodynamic arguments. Consider an incompressible solid with constant heat 
capacity $C$ (system $A$), which undergoes a process from an initial temperature $T+\Delta T$ 
due to thermal interaction with the environment at temperature $T$ (system $B$). The entropy 
variation of the solid is:
	\begin{equation}\label{DeltaS_A}
	\Delta S^{A}_{1\rightarrow 2}=C\log\left(\dfrac{T}{T+\Delta T}\right)=-C\log(1+x),
	\end{equation}
where $x=\Delta T/T$ (notice that  $x>-1$ since the environment cannot be at zero 
temperature). Applying the first law and using the fact that the variation of internal energy of an incompressible
solid is $\Delta U=C(T_f-T_i)$, we can obtain the energy exchanged by the environment during the process:
	\begin{equation}
	Q^{B}=-Q^{A}=C\Delta T.
	\end{equation}
Hence, the entropy variation of the environment (modeled as a thermal reservoir), is
	\begin{equation}\label{DeltaS_B}
	\Delta S^{B}_{1\rightarrow 2}=\dfrac{Q_{B}}{T}=Cx.
	\end{equation}
From Eqs. (\ref{DeltaS_A}) and (\ref{DeltaS_B}) we obtain:
	\begin{equation}\label{DeltaS_univ}
	\Delta S^{Univ}_{1\rightarrow 2}=\Delta S^{A}_{1\rightarrow 2}+\Delta S^{B}_{1\rightarrow 2}=C\left[x-\log(1+x)\right].
	\end{equation}
The key point is  that, according to the second law of thermodynamics, the global entropy change 
is always positive, so from Eq. (\ref{DeltaS_univ}) and using that $C>0$, we conclude that
	\begin{equation}
	\log(1+x)\leq x, \hspace{0.2cm} x>-1,
	\end{equation}
so the upper bound is established. On the other hand, the analysis of the 
opposite process due to thermal contact of the solid with a thermal reservoir $B'$ at the 
original temperature $T+\Delta T$ leads to the following entropy variations of the solid 
and the reservoir:
	\begin{equation}\label{DeltaS_A op}
	\Delta S^{A}_{2\rightarrow 1}=C\log\left(\dfrac{T+\Delta T}{T}\right)=C\log(1+x)
	\end{equation}
and
	\begin{equation}\label{DeltaS_B op}
	\Delta S^{B'}_{2\rightarrow 1}=\dfrac{Q_{B}'}{T+\Delta T}=\dfrac{-C\Delta T}{T+\Delta T}=\dfrac{-Cx}{1+x},
	\end{equation}
so from the global entropy balance we conclude that:
	\begin{equation}\label{DeltaS_univ op}
	\Delta S^{Univ}_{2\rightarrow 1}=C\left[\log(1+x)-\dfrac{x}{1+x}\right]\geq 0,
	\end{equation}
from which the lower bound is deduced.

The above procedure is completely general and allows us to obtain bounds for other types 
of functions if other systems and processes are 
considered instead of an incompressible solid. For example, consider the following 
bounds for the function $(1+x)^{\delta}$ (\textit{generalized Bernoulli's inequalities} \cite{Li}):
	\begin{equation}\label{Bernoulli}
	\dfrac{1+x}{1+x(1-\delta)}\leq(1+x)^{\delta}\leq 1+\delta x,
	\end{equation}
with $0<\delta <1$ and $x>-1$. To obtain these inequalities by thermodynamic arguments, consider 
an ideal 
gas with constant specific heat capacities $c_{P}$ and $c_{v}$ which is contained in a 
thermally isolated piston-cylinder device. The system is initially in an equilibrium state 
at temperature $T_{1}$ and pressure $P_{1}$ produced by the action of the atmosphere and 
the piston on the gas. Suppose that we suddenly add a load over the piston in such a way that
the external pressure increases to the value $P_{2}=P_{1}+\Delta P$.  It can be shown that 
after transient oscillations, the system will reach a new equilibrium state at pressure 
$P_{2}$ \cite{Mungan2017}. Define the quantities:
	\begin{equation}\label{xdelta}
	x=\dfrac{\Delta P}{P_1};\hspace{0.2cm} \delta =\dfrac{R}{c_{P}},
	\end{equation}
where $R$ is the gas constant. Performing the entropy analysis, the following inequality 
is obtained:

	\begin{equation}\label{Berneq}
	\log(1+\delta x)-\delta\log(1+x)\geq 0,
	\end{equation}
a result equivalent to the right inequality in Eq. (\ref{Bernoulli}). As the reader may 
suspect, the lower bound arises on considering the process that takes place when removing the 
load from the piston and waiting for the new equilibrium state to be reached at the initial 
pressure. The details of these derivations can be found in the Appendix B.

\section{Improving the bounds: general procedure}
Now consider a thermodynamic system (A) that undergoes an irreversible process $1\rightarrow 2$ 
between two equilibrium states due to  interaction with its environment (E). Since the global entropy 
change is non-negative:
	\begin{equation}\label{dSuniv}
	\Delta S^{Univ}=\Delta S^{A}+\Delta S^{E}\geq 0,
	\end{equation}
we must have
	\begin{equation}\label{bound12}
	-\Delta S^{E}\leq \Delta S^{A},
	\end{equation}
so the quantity $-\Delta S^{E}$ is a lower bound for $\Delta S^{A}$.

Now suppose that instead of performing the process $1\rightarrow 2$ directly, we have the system 
pass through an intermediate equilibrium state before reaching the final state. 
Note that, since entropy is a state function, the entropy variation of $A$ is the same in both cases.
Applying the second law to the global process, we obtain that
	\begin{equation}\label{dS''univ}
	\Delta S'^{Univ}=\Delta S^{A}+\Delta S'^{E}\geq 0,
	\end{equation}
where $\Delta S'^{Univ}$ and $\Delta S'^{E}$ denote the new entropy changes of the universe and the 
environment in the alternative process. Of course, this provides another lower bound for the entropy 
change of $A$:
	\begin{equation}\label{bound12'2}
	-\Delta S'^{E}\leq \Delta S^{A}.
	\end{equation}
Now, the inclusion of an intermediate equilibrium state in the trajectory 
implies that the new process is closer to the reversible limit than the original process and, as a 
consequence, 
less entropy is produced (see the Appendix A):
	\begin{equation}\label{dS'dS}
	\Delta S'^{Univ}\leq \Delta S^{Univ}.
	\end{equation}
From Eqs. (\ref{dSuniv}), (\ref{dS''univ}),  (\ref{bound12'2}) and (\ref{dS'dS}), we 
conclude that
	\begin{equation}
	-\Delta S^{E}\leq -\Delta S'^{E}\leq \Delta S^{A},
	\end{equation}
which implies that the lower bound for $\Delta S^{A}$ given by Eq. (\ref{bound12'2}) is sharper 
than that given by Eq. (\ref{bound12}). By including 
more intermediate equilibrium states in the process, sharper bounds for $\Delta S^{A}$ can be found.

\section{Application: Approximants for $\boldsymbol{\log(1+x)}$.}
\begin{figure}
		\centering
  		\includegraphics[trim= 15 0 0 0, scale=0.65, clip]{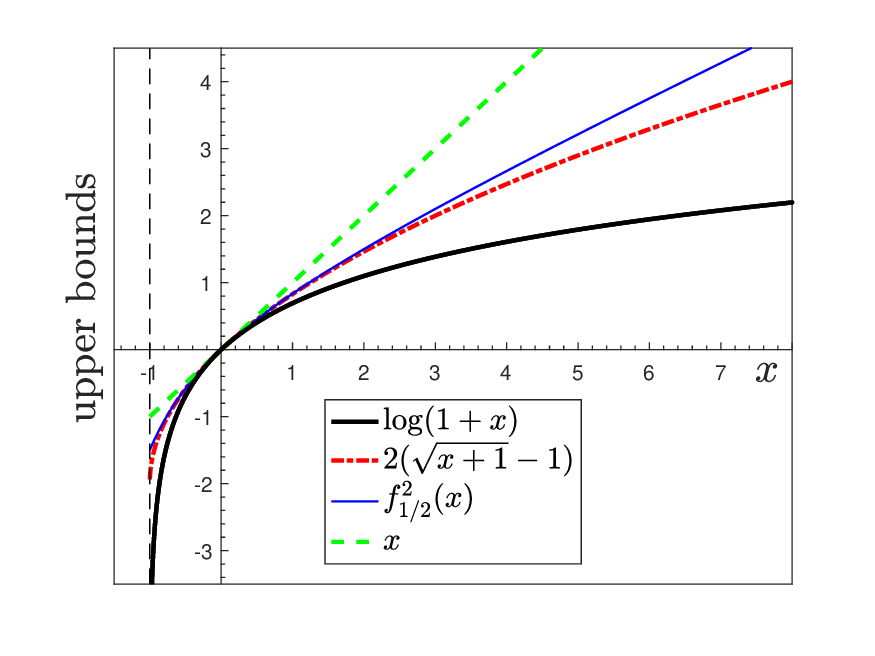}
		\caption{Plot of $f(x)=\log(1+x)$, its standard upper bound $h(x)=x$ and the majorizing 
		functions obtained dividing the process in two steps, for $p=1/2$.}
		\label{F1}
	\end{figure}

Here we apply the above method to the process that led to Eq. (\ref{logbounds}). As 
the method suggests, the idea is to perform the process from temperature $T+\Delta T$ 
to temperature $T$ in steps, introducing an intermediate equilibrium state. This can be done 
by placing the solid in contact with another thermal reservoir ($R$) at a temperature $T'=T+p\Delta T$, 
where $0\leq p\leq 1$. When thermal equilibrium with $R$ is reached 
(process $1\rightarrow 2'$), the energy exchanged by the reservoir $R$ is $Q^{R}=C(1-p)\Delta T$, 
so its entropy varies by the amount
	\begin{equation}\label{deltaSR}
	\Delta S^{R}_{1\rightarrow 2'}=\dfrac{Q^{R}}{T+p\Delta T}=\dfrac{C(1-p)x}{1+px}.
	\end{equation}
Finally, we remove the solid from contact with $R$ and let it release energy to the environment 
at temperature $T$ (process $2'\rightarrow 2$), so the final state of $A$ 
is the same as that in the original process. During the second stage, the energy absorbed by system 
$E$ is $Q^{E}=Cp\Delta T$, which implies an entropy increase of
	\begin{equation}\label{deltaSE}
	\Delta S^{E}_{2'\rightarrow 2}=\dfrac{Q^{E}}{T}=Cpx.
	\end{equation}
Then, from Eqs. (\ref{DeltaS_A}), (\ref{deltaSR}) and (\ref{deltaSE}), for the global process 
$1\rightarrow 2'\rightarrow 2$ we obtain 
	\begin{equation}
	\Delta S'^{Univ}_{1\rightarrow 2}=C\left[\dfrac{(1-p)x}{1+px}+px-\log(1+x)\right]\geq 0.
	\end{equation}
From this, the following family of upper bounds for $\log(1+x)$ arise:
	\begin{equation}\label{sharper_bound_1}
	\log(1+x)\leq\dfrac{x(1+p^2x)}{1+px}\equiv f^{2}_{p}(x),
	\end{equation}
for $x>-1,\hspace{0.2cm} 0\leq p\leq 1$. The superscript in $f^{2}_{p}(x)$ denotes the 
number of steps 
used, and the subscript represents the fraction of the global process covered at the end of each intermediate stage. 
Since the process $1\rightarrow 2'\rightarrow 2$ is closer to the reversible limit than  
original process, we conclude that (\ref{sharper_bound_1}) is a sharper upper bound for 
$\log(1+x)$ than that given by Eq. (\ref{logbounds}):
	\begin{equation}\label{logbounds11}
	\log(1+x)\leq f^{2}_{p}(x)\leq x,
	\end{equation}
with $x>-1,\hspace{0.2cm} 0\leq p\leq 1$; see Fig. (\ref{F1}). Taking the derivative of $f^{2}_{p}(x)$ 
with respect to $p$ and setting it equal to zero, we find that, for fixed $x$, 
the value of $p$ that minimizes the new bound is 
 $p_{min}=\dfrac{\sqrt{x+1}-1}{x}$, 
 which when replaced in Eq. (\ref{logbounds11}) allows us to conclude that
	\begin{equation}\label{logbounds2}
	\log(1+x)\leq 2(\sqrt{x+1}-1)\leq f^{2}_{p}(x)\leq x,
	\end{equation}
with $x>-1,\hspace{0.2cm} 0\leq p\leq 1$. This provides a potential approximant of order 1/2, 
certainly tighter than the family of approximations given by $f^{2}_{p}(x)$, whose growth 
for large values of $x$ is linear.

\begin{figure}
		\centering
  		\includegraphics[trim= 15 0 0 0, scale=0.65, clip]{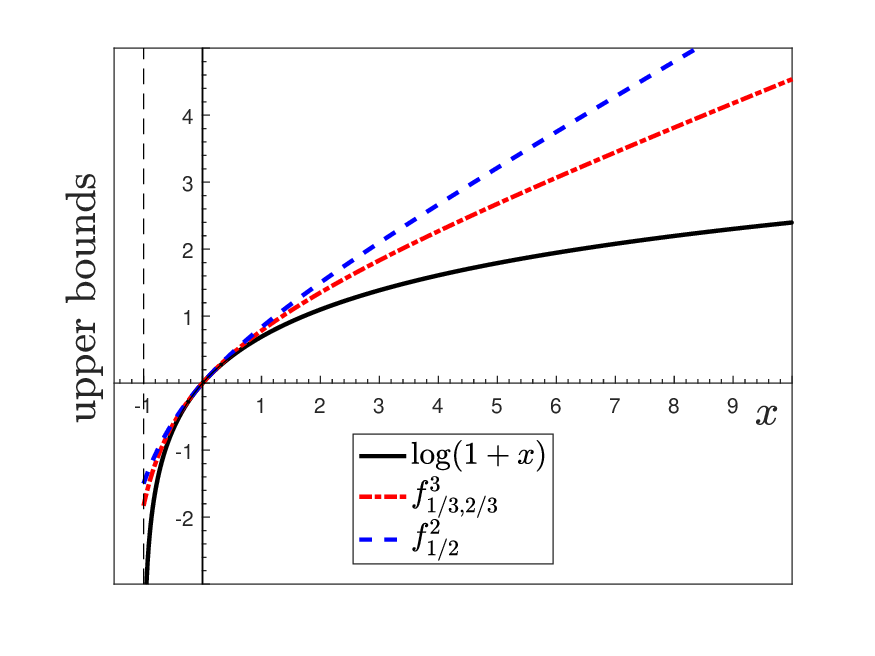}
		\caption{Comparison between the upper bounds for $\log(1+x)$ obtained partitioning the process into two and three steps. }
		\label{F2}
	\end{figure}

\begin{figure}
		\centering
		\includegraphics[trim= 15 0 0 0, scale=0.65, clip]{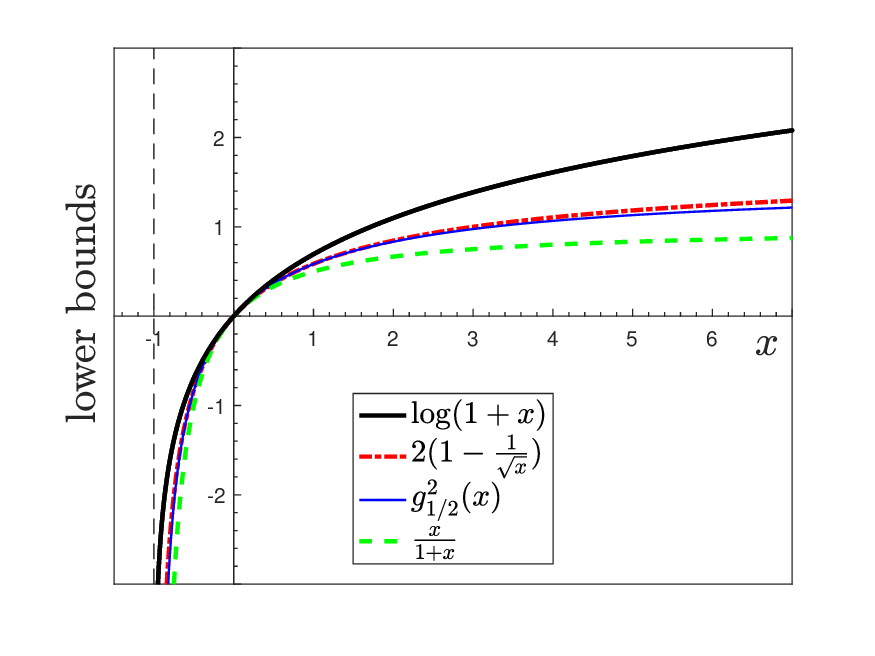}
		\caption{Plot of $f(x)=\log(1+x)$, its standard lower bound $j(x)=x/(1+x)$ and the minorizing functions 			obtained dividing the process in two steps, for $p=1/2$.}
		\label{F3}
	\end{figure}

As we have already mentioned, it is possible to obtain better bounds by dividing the process into 
more stages. As an example, let us assume that we perform the cooling process from 
temperature $T+\Delta T$ (state 1) to temperature $T $ (state 4) in three steps, allowing the 
system to thermalize with reservoirs at temperatures $T+2\Delta T/3$ (state 2), $T+\Delta T/3$ 
(state 3), and finally with the environment at temperature $T$. The total entropy production obtained in this process can be shown to be
	\begin{equation}
	\Delta S^{Univ}_{1\rightarrow 3}=C\left[x\left(\dfrac{1}{3+2x}+\dfrac{1}{3+x}+\dfrac{1}{3}\right)-\log(1+x)\right],
	\end{equation}
so comparing this process with the two-step cooling process already studied (using $p=1/2$), 
we obtain that
	\begin{equation}
	\log(1+x)\leq f^{3}_{1/3,2/3}\leq f^{2}_{1/2}\leq x,
	\end{equation}
for $x>-1$, where
	\begin{equation}
	f^{3}_{1/3,2/3}=\dfrac{x(2x^2+18x+27)}{3(3+x)(3+2x)},
	\end{equation}
and
	\begin{equation}
	f^{2}_{1/2}=\dfrac{x(4+x)}{4+2x}.
	\end{equation}
This is illustrated in Fig. (\ref{F2}).
Considering the heating process from temperature $T$ to temperature $T+\Delta T$ and following 
the same procedure, it is also possible to improve the lower-bound for $\log(1+x)$ given by Eq. 
(\ref{logbounds}). For instance, the reader may verify that if the system thermalizes with a 
reservoir at temperature $T+p\Delta T$ before reaching the final state, the following intermediate 
lower-bound can be deduced from the global entropy balance:
	\begin{equation}\label{lowerbounds}
	\dfrac{x}{1+x}\leq \dfrac{x[1+xp(2-p)]}{(1+x)(1+xp)}\equiv g^{2}_{p}(x)\leq\log(1+x),
	\end{equation}
with $x\geq -1$, and $0\leq p\leq 1$, see Fig. (\ref{F3}). The value of $p$ that maximizes $g^{2}_{p}(x)$ 
for each $x$ is the same that in the prevoius case, so substituting in (\ref{lowerbounds}), 
we obtain:
	\begin{equation}
	\dfrac{x}{1+x}\leq g^{2}_{p}(x)\leq 2\left(1-\dfrac{1}{\sqrt{x+1}}\right)\leq\log(1+x),
	\end{equation}
with $x\geq -1$, and $0\leq p\leq 1$.

\section{Discussion and conclusions}
The increase in entropy in every physical
process establishes a connection between thermodynamics and the theory 
of algebraic inequalities that, in our opinion, deserves further exploration. In the 
first part of this work we have employed this connection to obtain a well known set 
of bounds for the logarithmic function. The 
inequalities are being generated from a physics postulate, which is remarkable 
since they are true independent of any thermodynamic considerations (and even if the second 
law turns out to be false). Whether this kind of reasoning represents 
a formal proof of these inequalities has been the subject 
of debate in the past, particularly in the context of derivations of the inequality between 
the arithmetic and geometric means \cite{Abriata,Deakin,Sidhu}. Certainly, our view is critical 
about the logical consistency of 
thermodynamic proofs, since it is the maximization of entropy 
that assures that the imposed final state is, in fact, the equilibrium state. However, it is important 
to remark that we did not need to maximize the entropy to recognize 
the equilibrium state: it can be deduced from empirical considerations, such as the 
intuitive fact that the final temperature of a small system that exchanges energy with a 
thermal reservoir must coincide with the temperature 
of the reservoir. This implies that the underlying inequality associated with a particular 
process is never known or employed explicitly, but it is learned \textit{after} the entropy 
balance is performed. In that sense, it is undeniable that the thermodynamic method constitutes 
a credible path towards the construction of true algebraic inequalities. Some of them may be 
well-known results, but others are possibly new. This has an intrinsic value regardless 
of whether the process employed can or cannot be considered a formal proof. Of course, 
if the readers feel uncomfortable, they can try to demonstrate the inequalities that 
arise from the thermodynamic analysis through traditional methods.

It is a well-known fact that in partitioning a thermodynamic process to bring it closer to the 
reversible limit, less entropy is produced. This implies that when performing the entropy 
analysis for the new process, a different inequality is obtained in which the terms involved 
are closer than those generated 
in the original inequality. As we have shown, this property allows us to improve the 
bounds for the logarithmic function found in Section II, but, of course, the validity of 
the method is not restricted to that particular function. In fact, bounds for a wide class of 
functions can be obtained and improved if, instead of using ideal gases and incompressible solids, other 
substances (even hypothetical, with a more complex dependence of the entropy on the other thermodynamic
properties) are employed. 

We believe that, in addition to their power to deduce new algebraic inequalities, the methods 
applied in this work have an interesting didactic value and can motivate students, especially 
those with a markedly mathematical profile, to delve into the study of thermodynamics.

\section*{Acknowledgments}
This work was partially supported by Agencia Nacional de Investigaci\'on e Innovaci\'on (Uruguay). The author thanks the anonymous reviewers for helpful comments and suggestions.

\section*{Author Declarations}
The author has no conflicts to disclose.

\appendix
\section{Step processes and entropy production}
The fact that the introduction of an intermediate equilibrium state decreases the total entropy 
production has been theoretically and experimentally demonstrated in several contexts. For example, in Ref. \cite{Gupta} it is shown that 
if a linear spring is loaded to a total mass $M$ in $N$ equal steps, each time by a mass $M/N$, the total entropy production due to the energy dissipated to the environment by viscous friction is inversely proportional to the number of steps. The same dependence between the entropy production and N is observed in processes such as the charging of a capacitor, the compression of an ideal gas, or the heating of certain amount of water, when they are carried out in stages \cite{Gupta,Heinrich,Calkin}. In what follows we show that for the cooling process of an incompressible solid, such as the one studied in section IV, the introduction of intermediate reservoirs reduces the entropy generation. 

If the solid $A$ (initially at temperature $T_1$) is placed directly in thermal contact with a reservoir at temperature $T_N<T_1$,
the global entropy variation once the new equilibrium state is reached is given by
	\begin{equation}\label{dSU}
	\Delta S^{Univ}_{1\rightarrow N}=\Delta S^{A}_{1\rightarrow N}-\dfrac{\Delta U^{A}_{1\rightarrow N}}{T_N},
	\end{equation}
where $\Delta U^{A}_{1\rightarrow N}$ is the energy dissipated during the thermalization process.
On the other hand, if the system first thermalizes with a set of reservoirs at temperatures  $T_2,T_3,...T_{N-1}$ such that $T_1>T_2>...>T_{N-1}>T_{N}$, performing the entropy analysis for each process and adding the results we obtain that the new global entropy variation is:
	\begin{equation}\label{dSU'}
	\Delta S'^{Univ}_{1\rightarrow N}=\Delta S^{A}_{1\rightarrow N}-\sum_{j=1}^{N-1}\dfrac{\Delta U^{A}_{j		\rightarrow j+1}}{T_{j+1}}
	\end{equation}
From Eqs. (\ref{dSU}) and (\ref{dSU'}), and using that
	\begin{equation}
	\Delta U^{A}_{1\rightarrow N}=\sum_{j=1}^{N-1}\Delta U^{A}_{j\rightarrow j+1},
	\end{equation}
we obtain that the difference of the entropy variations is
	\begin{equation}\label{sum}
	\Delta S^{Univ}-\Delta S'^{Univ}=\sum_{j=1}^{N-2}\Delta U^{A}_{j\rightarrow j+1}\left(\dfrac{1}{T_{j+1}}-\dfrac{1}{T_N}\right).
	\end{equation}
Finally, noting that all energy variations $\Delta U^{A}_{j\rightarrow j+1}$ are negative (the solid releases energy) and that $T_N<T_{j+1}$ for all $j=1,...,N-1$, we have that each term in the right-hand side of Eq. (\ref{sum}) must be positive, so we conclude that
	\begin{equation}
	\Delta S'^{Univ}<\Delta S^{Univ},
	\end{equation}
and hence the new process must be closer to the reversible limit than the original.

\section{Thermodynamic derivation of the generalized Bernoulli's inequality}

Let us return to the analysis of the process suffered by a perfect gas contained in an adiabatic cylinder-piston device when a mass is suddenly added over the piston. It is important to remark that if the mass is not infinitesimal, the process followed by the gas will be non-quasi-static, and the internal pressure may be not be well-defined during the process. Nevertheless, since the external pressure is constant, the work performed by the gas can be calculated as \cite{Gislason}:
	\begin{equation}\label{W1}
	W=\int_{V1}^{V2}P_{ext}dV=P_{2}(V_{2}-V_{1}).
	\end{equation}
The initial and final volumes can be expressed in terms of the pressures and temperatures using the equation of state for an ideal gas:
	\begin{equation}\label{deltaV}
	V_{2}-V_{1}=R\left[\dfrac{T_{2}}{P_{2}}-\dfrac{T_{1}}{P_{1}}\right].
	\end{equation}
Substituting Eqs. (\ref{W1}) and (\ref{deltaV}) into the first law for an adiabatic process:
	\begin{equation}\label{1law}
	\Delta U=-W,
	\end{equation}
and using that the internal energy is a function of the temperature only \cite{Cengel}:
	\begin{equation}\label{DeltaU}
	\Delta U =c_{v}\Delta T,
	\end{equation}
we obtain that
	\begin{equation}
	c_{v}\left(T_{2}-T_{1}\right)=-R\left[T_{2}-\dfrac{P_{2}T_{1}}{P_{1}}\right].
	\end{equation}  
From the above equation, and using that for an ideal gas $c_{p}-c_{v}=R$, we find the temperature of the final state:
	\begin{equation}
	T_{2}=\left[\dfrac{c_{v}}{c_{p}}+\dfrac{R{P_{2}}}{c_{p}P_{1}}\right]T_{1},
	\end{equation}
\noindent which can be expressed as:
	\begin{equation}\label{T2}
	T_{2}=T_{1}\left(1+\delta x\right),
	\end{equation}
\noindent where $x$ and $\delta$ are given by Eq. (\ref{xdelta}).

Now we are in position to analyze the process from the perspective of the second law. Since the entropy of the environment does not change, the total entropy production coincides with the entropy variation of the gas, which can be evaluated from the equation:
 
	\begin{equation}\label{DeltaSgas}
	\Delta S^{gas}_{1\rightarrow 2}=c_{P}\log\left(\dfrac{T_{2}}{T_{1}}\right)-R\log\left(\dfrac{P_{2}}{P_{1}}\right),
	\end{equation}
which, combined with Eq. (\ref{T2}) and the relation $P_{2}/P_{1}=1+x$, gives
	\begin{equation}\label{Sgen1}
	\dfrac{\Delta S^{Univ}_{1\rightarrow 2}}{c_{p}}=\log(1+\delta x)-\delta\log(1+x)\geq 0,
	\end{equation}
from which the upper bound
	\begin{equation}
	(1+x)^{\delta}\leq 1+\delta x
	\end{equation}
is obtained. It is easy to verify that the conditions $x>-1$ and $0<\delta <1$ are satisfied 
on noting that $P$ is a non-negative quantity and that $R<c_P$ for an ideal gas. However, for the proof 
to be valid for any values of $x$ and $\delta $ in the described ranges, we must assume the 
existence of perfect 
gases with all values of $\delta$ in $(0,1)$, and that the gas is still ideal even for 
large values of pressure, situation in which the hypothesis that the molecules are not-interacting 
ceases to be reasonable.  

Regarding the lower bound, it can be found by analyzing the process $2\rightarrow 3$ undergone 
by the gas after removing the load from the piston. Following the same procedure, we obtain that 
the new equilibrium state is defined by the following properties:
	\begin{equation}
	P_{3}=P_{1}, \hspace{0.2cm} T_{3}=T_{2}\left[\dfrac{1+x(1-\delta)}{1+x}\right]
	\end{equation}
so the entropy created during the expansion satisfies

	\begin{equation}\label{DS_univ_exp2}
	\dfrac{\Delta S^{Univ}_{2\rightarrow 3}}{C_{p}}=\log\left[\dfrac{1+x(1-\delta)}{1+x}\right]+\delta\log(1+x)\geq 0,
	\end{equation}
a result that proves the left inequality in Eq. (\ref{Bernoulli}).

\end{document}